\documentclass[12pt,twoside]{article}
\usepackage[mathscr]{eucal}
\usepackage{amsmath,amsfonts,amssymb,amsthm,mathabx,empheq}
\usepackage{times}
\usepackage{pdfsync}
\usepackage{cite}
\usepackage{url}
\usepackage{hyperref}
\usepackage{tensor}
\usepackage{color}
\usepackage{multicol}
\usepackage{bbold}

\advance\textheight\topskip
\allowdisplaybreaks[1]

\newcommand\td{\text{d}}

\newcommand{\p}{\partial}

\newcommand{\be}{\begin{equation}}
\newcommand{\ee}{\end{equation}}
\newcommand{\bea}{\begin{eqnarray}}
\newcommand{\eea}{\end{eqnarray}}

\def\n{\nabla}

\newcommand{\nn}{\nonumber}
\newcommand*\xbar[1]{%
  \hbox{%
    \vbox{%
      \hrule height 0.5pt 
      \kern0.3ex
      \hbox{%
        \kern-0.0em
        \ensuremath{#1}%
        \kern-0.0em
      }%
    }%
  }%
}

\allowdisplaybreaks[1]

\DeclareFontFamily{OT1}{rsfs}{} \DeclareFontShape{OT1}{rsfs}{m}{n}{
<-7> rsfs5 <7-10> rsfs7 <10-> rsfs10}{}
\DeclareMathAlphabet{\mycal}{OT1}{rsfs}{m}{n}

\hypersetup{
colorlinks=true,
linktoc=page,    
linkcolor=blue,
citecolor=blue,
urlcolor=blue,
colorlinks=true,
}

\begin{document}

\title{Weyl double copy in type D spacetime in four and five dimensions}

\author{Weicheng Zhao, Pu-Jian Mao, and Jun-Bao Wu}

\date{}

\def\mytitle{Weyl double copy in type D spacetime in four and five dimensions}

\begin{flushright}
\tt USTC-ICTS/PCFT-24-42
\end{flushright}

\addtolength{\headsep}{4pt}

\begin{centering}

  \vspace{1cm}

  \textbf{\large{\mytitle}}

  \vspace{1.5cm}

  {\large Weicheng Zhao$^a\,$\footnote{
The unusual ordering of authors instead of the standard alphabetical one in hep-th community is to ensure that the student receives proper recognition for the student's contribution under the current outdated practices in China.
}, Pu-Jian Mao$^{a}$, and Jun-Bao Wu$^{a,b}$ }

\vspace{0.5cm}

\begin{minipage}{.9\textwidth}\small \it  \begin{center}
     ${}^{a}$ Center for Joint Quantum Studies and Department of Physics,\\
     School of Science, Tianjin University, 135 Yaguan Road, Tianjin 300350, China
 \end{center}
\end{minipage}

\vspace{0.3cm}

\begin{minipage}{.9\textwidth}\small \it  \begin{center}
    ${}^{b}$ Peng Huanwu Center for Fundamental Theory,  Hefei, Anhui 230026, China
 \end{center}
 \end{minipage}

\vspace{0.3cm}

\end{centering}

\begin{center}
Emails: zhaoweichengok@tju.edu.cn,\, pjmao@tju.edu.cn,\, junbao.wu@tju.edu.cn
\end{center}

\begin{center}
\begin{minipage}{.9\textwidth}
\textsc{Abstract}: In this paper, we present a second realization of the Weyl double copy (WDC) in four-dimensional algebraic type D spacetime. We show that any type D vacuum solution admits an algebraically general Maxwell scalar on the curved background that squares to give the Weyl scalar. The WDC relation defines a scalar field that satisfies the Klein-Gordon equation sourced by the Weyl scalar on the curved background. We then extend the type D WDC to five dimensions.

\end{minipage}
\end{center}

\section{Introduction}

The double copy relation reveals remarkable connections between gauge and gravity theories, which indicates the perturbative scattering amplitude of gravity as a product of two scattering amplitudes of gauge theory \cite{Kawai:1985xq,Bern:2008qj,Bern:2010ue}. Recently, the double copy relation is shown to also exist at the classical level connecting classical solutions of gauge and gravity theories. The first such example is constructed in the Kerr-Schild form \cite{Monteiro:2014cda}, which reveals exact relations between solutions beyond perturbative approach. The classical double copy underpins the intrinsic connection between the gauge and gravity theories and significantly enlarges the scope of the double copy relation outside of the amplitudes community, see, e.g., \cite{Luna:2015paa,Luna:2016due,Goldberger:2016iau,Goldberger:2017frp,Bahjat-Abbas:2017htu,Carrillo-Gonzalez:2017iyj,Shen:2018ebu,Lee:2018gxc,Berman:2018hwd,CarrilloGonzalez:2019gof,Lescano:2020nve,Easson:2020esh,Gumus:2020hbb,Almeida:2020mrg,Lescano:2021ooe,Campiglia:2021srh,Alkac:2021bav,Alkac:2021seh,Chacon:2021hfe,Adamo:2021dfg,Mao:2021kxq,Shi:2021qsb,Didenko:2022qxq,Chawla:2023bsu,Ortaggio:2023rzp,Ortaggio:2023cdz,Ceresole:2023wxg,Ferrero:2024eva}. Shortly, the classical double copy relation is realized in a gauge invariant way by connecting the Weyl tensor of the gravity theory and the field strength tensor of the Maxwell theory \cite{Luna:2018dpt}, which is known as Weyl double copy (WDC). The WDC provides a coordinate independent framework to investigate the exact mathematical connections between Einstein and Maxwell's theories, which is now  at the center of attention of many researchers.

The WDC is better appreciated in the spinorial formalism \cite{Luna:2018dpt,Godazgar:2020zbv} and has clear  twistorial foundation \cite{White:2020sfn,Chacon:2021wbr}. Hence, the recent developments are mainly limited in four-dimensional spacetime \cite{Luna:2018dpt,Keeler:2020rcv,Alawadhi:2020jrv,Godazgar:2020zbv,White:2020sfn,Monteiro:2020plf,Chacon:2021wbr,Godazgar:2021iae,Easson:2021asd,Chacon:2021lox,Han:2022ubu,Han:2022mze,Luna:2022dxo,Chawla:2022ogv,Easson:2022zoh,Easson:2023dbk,Alkac:2023glx,Mao:2023yle,Liu:2024byr,Chawla:2024mse,Armstrong-Williams:2024bog}.\footnote{In principle, the WDC relation in \cite{Luna:2018dpt} can be applied for particular exact solutions in higher dimensions if there is a spinorial description of the relevant fields. The single copy of the Kerr-NUT-(A)dS spacetime in five dimensions was derived from the WDC relation \cite{Chawla:2022ogv}. The purely
algebraic Weyl doubling constraints to the Weyl tensor was examined in \cite{Alawadhi:2020jrv} in generic dimensions, see also \cite{Didenko:2011ir} for earlier relevant investigations in this direction. In three-dimensional spacetime, a Cotton double copy was proposed in \cite{CarrilloGonzalez:2022mxx,CarrilloGonzalez:2022ggn}.} In a previous work \cite{Zhao:2024ljb}, we generalized the WDC to five dimensions for a special class of type N solutions in the Coley-Milson-Pravda-Pravdova (CMPP) classification \cite{Coley:2004jv,Milson:2004jx,Pravda:2004ka,Ortaggio:2012jd}. The aim of the present paper is to extend the previous investigation \cite{Zhao:2024ljb} to five-dimensional type D solutions. The particular motivation to establish a type D WDC is that it can offer a double copy perspective for black hole solutions which would significantly extend the validity of the WDC in five dimensions. Considering two equal electromagnetic fields, the four-dimensional WDC formula is only valid for the algebraic type D and type N spacetime.\footnote{For different electromagnetic fields, the WDC relation can be defined in other types of spacetime \cite{White:2020sfn}.} However, the realizations of the WDC for type D and type N spacetime are very different. For the type N case \cite{Godazgar:2020zbv}, the WDC is realized at the equations of motion level. It is proven from the Bianchi identities of the Weyl tensor and the Maxwell's equations that any type
N vacuum solution admits a degenerate Maxwell field that squares to give the Weyl tensor and the WDC relation defines a scalar field that satisfies the source-free Klein-Gordon equation on the curved background.  For the type D case \cite{Luna:2018dpt}, the WDC is proposed and tested for the Plebanski-Demianski solution, the most general known type D solution \cite{Plebanski:1976gy}. The Maxwell field that squares to give the Weyl tensor is taken from the Kerr-Schild double copy relation and a scalar field is defined from the WDC relation which satisfies the wave equation on the Minkowski background. The main difficulty for generalizing the four-dimensional type D WDC to five dimensions is that there is no generic solution which can represent the type D spacetime in five dimensions. Hence, a crucial step for the generalization is to realize the WDC at the equations of motion level which is the first goal of this work.

In this paper, we will present a second realization of the type D WDC in the Newman-Penrose (NP) formalism \cite{Newman:1961qr}. We show at the equations of motion level that the Weyl scalar of any type D solution can be decomposed as a product of two algebraically general Maxwell scalars and one scalar field which satisfies the wave equation sourced by the
Weyl scalar on the type D background. We then extend the type D WDC to five dimensions in the vielbein formalism. We find a self-contained reduction of the type D spacetime in the CMPP classification, where one can confirm that any solution of this special class admits a special class of algebraically general Maxwell field that squares to give the Weyl tensor. Moreover, a complex scalar field is defined from the WDC relation which solves the wave equation sourced by the Weyl scalar and a special spin coefficient on the type D background. Hence, a concrete type D WDC relation is uncovered in five dimensions. Our results provide another example of the higher-dimensional WDC, which reveals the robustness of the classical double copy relation.

\section{New realization of the WDC for 4d type D spacetime}

We revisit the type D WDC relation in the NP formalism in this section. We will follow the conventions in \cite{Chandrasekhar} for the NP formalism. For a type D spacetime, one can arrange that the null bases $l$ and $n$ are both repeated principle null directions. Then the only non-zero Weyl scalar is $\Psi_2$. The Goldberg–Sachs theorem \cite{Goldberg} indicates that the spin coefficients $\kappa=\sigma=\lambda=\nu=0$. One can further use a third class rotation of the null bases to set $\epsilon=0$. The Bianchi identities for the Weyl scalar now become
\begin{equation}\label{4dbianchi}
        D\Psi_2=3\rho\Psi_2,\quad\quad \Delta\Psi_2=-3\mu\Psi_2,\quad \quad 
        \delta\Psi_2=3\tau\Psi_2,\quad\quad \bar{\delta}\Psi_2=-3\pi\Psi_2.
\end{equation}
We introduce an algebraically general Maxwell field which is arranged as that the only non-zero Maxwell scalar is $\Phi_1$. Then, the Maxwell's equations on the type D background are reduced to
\begin{equation}\label{4dmaxwell}
        D\Phi_1=2\rho\Phi_1,\quad\quad
        \Delta\Phi_1=-2\mu\Phi_1,\quad\quad
        \delta\Phi_1=2\tau\Phi_1,\quad\quad
        \bar{\delta}\Phi_1=-2\pi\Phi_1.
\end{equation}
The type D WDC relation in the NP formalism is given by \cite{Godazgar:2021iae}
\begin{equation}\label{4dwdc}
    \Psi_2=\frac{1}{S}\Phi_1^2.
\end{equation}
Substituting the WDC formula \eqref{4dwdc} into the Bianchi identities and applying the Maxwell's equations, we obtain the equations for the scalar field $S$
\begin{equation}\label{4dscalar}
        D\log{S}=\rho,\quad\quad
        \Delta\log{S}=-\mu,\quad\quad
        \delta\log{S}=\tau,\quad\quad
        \bar{\delta}\log{S}=-\pi.
\end{equation}
Clearly, any solution of the equations in \eqref{4dscalar} for the scalar field $S$ will yield a Maxwell scalar from the square-root relation 
\be
\Phi_1=\sqrt{S \Psi_2}.
\ee
One can prove that the equations in \eqref{4dscalar} satisfy the integrability conditions
\begin{equation}
\begin{split}
    &(\delta D-D\delta)\log{S}=\delta\rho-D\tau,\quad (\bar{\delta}\Delta-\Delta\bar{\delta})\log{S}=\Delta\pi-\bar{\delta}\mu,\\
    &(\Delta D-D\Delta)\log{S}=\Delta\rho+D\mu,\quad (\bar{\delta}\delta-\delta\bar{\delta})\log{S}=\bar{\delta}\tau+\delta\pi,\\
    &(\bar{\delta}D-D\bar{\delta})\log{S}=\bar{\delta}\rho+D\pi,\quad (\delta\Delta-\Delta\delta)\log{S}=-\delta\mu-\Delta\tau.
    \end{split}
\end{equation}
The integrability conditions in the first line can be easily proven using the commutators in the Appendix \ref{4dcommutator} and the Ricci identities in the Appendix \ref{4dricci}. The integrability conditions in the last line are equivalent to the following two equations
\begin{equation}
    \begin{split}
    &\bar{\delta}\rho+D\pi=\rho(\alpha+\bar{\beta}),\\
    &\delta\mu+\Delta\tau=-\mu(\bar{\alpha}+\beta)+\tau(\gamma-\bar{\gamma}),
    \end{split}
\end{equation}
which are nothing but the two identities obtained by acting the commutators $\bar{\delta}D-D\bar{\delta}$ and $\delta\Delta-\Delta\delta$ on $\Psi_2$ and simplifying the left hand side of the commutators by the Bianchi identities \cite{Kinnersley:1969zza}. The integrability conditions in the second line are equivalent to the following equations
\begin{equation}\label{identity}
    \begin{split}
    &\Delta\rho+D\mu=\rho(\gamma+\bar{\gamma})-\tau\bar{\tau}+\pi\bar{\pi},\\
    &\bar{\delta}\tau+\delta\pi=\bar{\mu}\rho-\bar{\rho}\mu+(\alpha-\bar{\beta})\tau+(\bar{\alpha}-\beta)\pi.
    \end{split}
\end{equation}
The first equation is nothing but the identity obtained by acting the commutator $\Delta D-D\Delta$ on $\Psi_2$ and simplifying the left hand side of the commutator by the Bianchi identities \cite{Kinnersley:1969zza}. The second one can be derived from the combination of the following two Ricci identities
\begin{equation}
    \begin{split}
        &\Delta\rho-\bar{\delta}\tau=-\rho\bar{\mu}+\tau(\bar{\beta}-\alpha-\bar{\tau})+\rho(\gamma+\bar{\gamma})-\Psi_2,\\
        &D\mu-\delta\pi=\bar{\rho}\mu+\pi(\bar{\pi}+\beta-\bar{\alpha})+\Psi_2,
    \end{split}
\end{equation}
and the first identity in \eqref{identity}. We hence verify that there must be solutions to the equations in \eqref{4dscalar} for the scalar field $S$, which guarantees that any type D vacuum solution admits an algebraically general Maxwell tensor on the type D background that squares to give the Weyl tensor. One can show that any solution to the equations in \eqref{4dscalar} must satisfy the following wave equation
\begin{equation}
    (\Box + 2 \Psi_2 )S=0.
\end{equation}
This wave equation is consistent with \cite{Godazgar:2021iae}. The Weyl scalar $\Psi_2$ is the source term in this wave equation, which is different from the type N case where the scalar field satisfies source-free wave equation. Nevertheless, it is reasonable to include a source term for the type D spacetime since it usually describes black hole solution. The source term somehow allows that the wave equation can be even valid at the singularity of the black hole. For the case of the Plebanski-Demianski solution, the scalar field also satisfies the wave equation on the flat background \cite{Luna:2018dpt} which simply indicates that the difference of the Box operator $\Box$ in the curved type D spacetime and the flat Minkowski spacetime is compensated by the extra source term $\Psi_2$.

Before closing this section, we will comment on the similarity of the Bianchi identities \eqref{4dbianchi}, the Maxwell's equations \eqref{4dmaxwell}, and the scalar equations \eqref{4dscalar}. Essentially, they are the same equations when the fields are properly redefined, e.g., imposing $S^3=( \Phi_1)^2=\Psi_2$. In this sense, our realization seems fake and can not provide any new information than testing the generic solutions \cite{Luna:2018dpt}. The integrability conditions we have proven simply indicate the existence of type D solutions. But the point of the new realization is for the extension to five dimensions. We will see in the next sections that the similarity between the three classes of equations disappears in five dimensions. But the verification of the WDC can be repeated in five dimensions for a special class of type D vacuum solutions.

\section{5d reduced type D spacetime}

In this section, we will specify a special class of type D spacetime in the CMPP classification in five dimensions, which is based on the following vielbein bases
\begin{equation}
    l=e_{0}=e^{1}, \quad n=e_{1}=e^{0},\quad m_{i}=e_{i}=m^{i}=e^{i},\quad i=2,3,4,
\end{equation}
where $l$ and $n$ are null, and $m_i$ are spacelike. The basis vectors satisfy the orthogonal conditions
\begin{equation}
    l \cdot l =n \cdot n= l \cdot m^{i}= n\cdot m^{i}=0,
\end{equation}
and the normalization conditions
\begin{equation}
    l\cdot n=1,\quad m^{i} \cdot m^{j}=\delta^{ij}.
\end{equation}
As directional derivatives, the basis vectors are assigned with special symbols
\be
D=l^\mu\p_\mu,\quad \quad \Delta=n^\mu\p_\mu,\quad\quad \delta_i=m_i^\mu\p_\mu.
\ee
The following quantities are defined to denote the spin coefficients
\be
L_{\mu\nu}= \n_\nu l_\mu ,\quad N_{\mu\nu}= \n_\nu n_\mu,\quad  M_{\mu\nu}^k=\n_\nu  m^{k}_\mu.
\ee
The CMPP classification is based on the maximum boost order of tensors which depends only on the choice of the null direction $l$ \cite{Coley:2004jv,Milson:2004jx,Pravda:2004ka,Ortaggio:2012jd}. The Weyl tensor can be decomposed as
\begin{align}
&C_{\mu\nu\rho\sigma} = \ \underbrace{4C_{0i0j} n_{\{\mu} m_{\nu}^{i} n_{\rho} m_{\sigma\}}^{j}}_{\text{boost weight } +2}
+ \underbrace{8C_{010i} n_{\{\mu} l_{\nu} n_{\rho} m_{\sigma\}}^{i}+4C_{0ijk} n_{\{\mu} m_{\nu}^{i} m_{\rho}^{j} m_{\sigma\}}^{k}}_{+1}\nn\\
&+ \underbrace{4C_{0101} n_{\{\mu} l_{\nu} n_{\rho} l_{\sigma\}}
+4C_{01ij} n_{\{\mu} l_{\nu} m_{\rho}^{i} m_{\sigma\}}^{j}
+8C_{0i1j} n_{\{\mu} m_{\nu}^{i} l_{\rho} m_{\sigma\}}^{j}
+ C_{ijkl} m_{\{\mu}^{i} m_{\nu}^{j} m_{\rho}^{k} m_{\sigma\}}^{l}}_{0}\nn\\
&\hspace{2cm} + \underbrace{8C_{101i} l_{\{\mu} n_{\nu} l_{\rho} m_{\sigma\}}^{i}
+ 4C_{1ijk} l_{\{\mu} m_{\nu}^{i} m_{\rho}^{j} m_{\sigma\}}^{k}}_{-1}
+ \underbrace{4C_{1i1j} l_{\{\mu} m_{\nu}^{i} l_{\rho} m_{\sigma\}}^{j}}_{-2}.\label{typed}
\end{align}
where we use the notation $C_{\mu\nu\rho\sigma}=C_{\{\mu\nu\rho\sigma\}}\equiv\frac{1}{2}(C_{[\mu\nu][\rho\sigma]}+C_{[\rho\sigma][\mu\nu]})
$. For a type D spacetime, the non-zero components of the Weyl tensor must be of boost weight zero. Hence, one has 
\be
C_{0i0j}=C_{010i}=C_{0ijk}=C_{101i}=C_{1ijk}=C_{1i1j}=0,
\ee
for the type D spacetime. Though it is algebraically special, there are still too many non-zero Weyl scalars to reveal a WDC in five dimensions. We will introduce a reduction of the five-dimensional type D spacetime for realizing a concrete WDC. The precise reductions of the five-dimensional type D spacetime in the vielbein formalism are
\begin{equation}
    \begin{split}
        &C_{\hat{a}\hat{b}\hat{c}4}=0,\quad\quad \hat{a}=0,1,...,4,\\
        &L_{i0}=L_{10}=M^i_{j0}=0.
    \end{split}
\end{equation}
Combining the type D condition, the above reductions, and the special arrangements of the spin coefficients according to the five dimensional Goldberg-Sachs-like theorem \cite{Durkee:2009nm,Ortaggio:2012hc}, we obtain the following constraints
\begin{equation}
    \begin{split}
        &C_{0331}=C_{0221},\quad C_{0321}=-C_{0231},\\
        &C_{0123}=-2C_{0231},\quad C_{0101}=-2C_{0221},\quad C_{2323}=2C_{0221},\\
        &L_{10}=N_{00}=L_{i0}=N_{i1}=0,\quad M^i_{j0}=0,\\
        &L_{4i}=L_{i4}=N_{4i}=N_{i4}=M^i_{41}=M^i_{44}=0,\quad M^3_{42}=M^2_{43}=0,\\
        &L_{41}=N_{40}=M^2_{42}=M^3_{43},\\
        &L_{22}=L_{33},\quad L_{32}=-L_{23},\quad N_{22}=N_{33},\quad N_{32}=-N_{23},
           \end{split}
\end{equation}
where the derivation of those conditions is detailed in the Appendix \ref{5dreduction}. Correspondingly, the Maxwell scalars are constrained as
\begin{equation}
    F_{\hat{a}4}=0,\quad F_{0i}=F_{1i}=0.
\end{equation}
We introduce the following definitions
\begin{equation}
\begin{split}
     &\psi_2=-C_{0221}-iC_{0231},\quad 
 \phi_1=\frac{1}{2}(F_{01}+i F_{23}),\\
 &\delta=\frac{1}{\sqrt{2}}(\delta_2-i\delta_3),\quad \bar{\delta}=\frac{1}{\sqrt{2}}(\delta_2+i\delta_3), \quad \delta'=i\delta_4.
\end{split}
\end{equation}
Then, the Bianchi identities in the Appendix \ref{Bianchi5d} can be rewritten as
\begin{equation}\label{5dbianchi}
    \begin{split}
        &D\psi_2=-3(L_{22}+iL_{23})\psi_2,\quad\quad
        \Delta \psi_2=-3(N_{22}-iN_{23})\psi_2,\\
        &\delta\psi_2=\frac{3}{\sqrt{2}}(L_{21}-iL_{31})\psi_2,\quad\quad
        \bar{\delta} \psi_2=\frac{3}{\sqrt{2}}(N_{20}+iN_{30})\psi_2,\\
        &\delta'\psi_2=2iL_{41}\psi_2.
    \end{split}
\end{equation}
The Maxwell's equations in the Appendix \ref{Maxwell5d} are reorganized as
\begin{equation}\label{5dmaxwell}
    \begin{split}
        &D\phi_1=-2(L_{22}+iL_{23})\phi_1,\quad\quad
        \Delta \phi_1=-2(N_{22}-iN_{23})\phi_1,\\
        &\delta \phi_1=\frac{2}{\sqrt{2}}(L_{21}-iL_{31})\phi_1,\quad\quad
        \bar{\delta}\phi_1=\frac{2}{\sqrt{2}}(N_{20}+iN_{30})\phi_1,\\
        &\delta'\phi_1=2iL_{41}\phi_1.
    \end{split}
\end{equation}
In four dimensions, the Bianchi identities \eqref{4dbianchi} and the Maxwell's equations \eqref{4dmaxwell} are aligned. In five dimensions, the fifth derivative $\delta'$ breaks the alignment, which challenges the extension of the WDC relation to five dimensions. Nevertheless, we will prove in the next section that a five-dimensional WDC can be realized in this reduced  type D case.

\section{WDC for 5d type D spacetime}

With the ingredients presented in the previous section, we propose that the five-dimensional WDC relation for the reduced type D spacetime should be
\begin{equation}
    \psi_2=\frac{1}{S}(\phi_1)^2
\end{equation}
The WDC relation yields the equations for the scalar field as
\begin{equation}\label{5dscalar}
    \begin{split}
        &D\log{S}=-(L_{22}+iL_{23}),\quad\quad
        \Delta \log{S}=-(N_{22}-iN_{23}),\\
        &\delta \log{S}=\frac{1}{\sqrt{2}}(L_{21}-iL_{31}),\quad\quad
        \bar{\delta}\log{S}=\frac{1}{\sqrt{2}}(N_{20}+iN_{30}),\\
        &\delta'\log{S}=2iL_{41}.
    \end{split}
\end{equation}
Again, the fifth derivative $\delta'$ breaks the  alignment of the scalar equations to the Bianchi identities and the Maxwell's equations in five dimensions.

Using the commutators in the Appendix \ref{commutator5d} and the Ricci identities in the Appendix \ref{ricci5d} (see \cite{Ortaggio:2007eg} for the full Ricci identities) one can easily prove the following four integrability conditions for the scalar field
\begin{equation}
    \begin{split}
        &(\delta D-D \delta) \log{S}=-\delta(L_{22}+iL_{23})-\frac{1}{\sqrt{2}}D (L_{21}-iL_{31}),\\
        &(\bar{\delta}\Delta-\Delta \bar{\delta})\log{S}=-\bar{\delta}(N_{22}-iN_{23})-\frac{1}{\sqrt{2}}\Delta (N_{20}+iN_{30}),\\
        &(\delta'D-D\delta')\log{S}=-\delta'(L_{22}+iL_{23}) - 2iD L_{41},\\
        &(\delta \delta'-\delta' \delta)\log{S}=2iD \delta L_{41} - \frac{1}{\sqrt{2}}\delta' (L_{21}-iL_{31}).
    \end{split}
\end{equation}
Using the Ricci identities, one can show that the following two integrability conditions
\begin{equation}
\begin{split}
    &(\Delta D-D\Delta) \log{S}=-\Delta (L_{22}+iL_{23}) + D (N_{22}-iN_{23}),\\
    &(\bar{\delta}\delta-\delta \bar{\delta})\log{S}= \frac{1}{\sqrt{2}}\bar{\delta} (L_{21}-iL_{31}) - \frac{1}{\sqrt{2}}\delta(N_{20}+iN_{30}),
\end{split}
\end{equation}
are equivalent to the same relation
\begin{equation}\label{5}
    D(N_{22}-iN_{23})-\Delta(L_{22}+iL_{23})=-L_{11}(L_{22}+iL_{23})+\frac{1}{2}[(L_{21})^2+(L_{31})^2-(N_{20})^2-(N_{30})^2],
\end{equation}
which is nothing but an identity similar to the four-dimensional analysis. More precisely, one can act the commutator $\Delta D-D\Delta$ on $\psi_2$ and apply once the Bianchi identities to yield
\be
(\Delta D-D\Delta )\psi_2=3\psi_2\left[-L_{11}(L_{22}+iL_{23})+\frac{1}{2}\left((L_{21})^2+(L_{31})^2-(N_{20})^2-(N_{30})^2\right)\right].
\ee
The same commutator relation can also be computed directly by using twice the Bianchi identities as 
\be
\Delta(D\psi_2)-D(\Delta\psi_2)=3\psi_2 D(N_{22}-iN_{23})-3\psi_2\Delta(L_{22}+iL_{23}).
\ee
The equivalence of the two equations verifies the relation \eqref{5}. The following two integrability conditions
\begin{equation}
    \begin{split}
        &(\bar{\delta}D-D\bar{\delta})\log{S}=-\bar{\delta}(L_{22}+iL_{23})- \frac{1}{\sqrt{2}}D (N_{20}+iN_{30}),\\
        &(\delta \Delta-\Delta \delta)\log{S}=-\delta(N_{22}-iN_{23})-\frac{1}{\sqrt{2}}\Delta(L_{21}-iL_{31}).
    \end{split}
\end{equation}
can be verified similarly from the actions of the commutators $\bar{\delta}D-D\bar{\delta}$ and $\delta \Delta-\Delta \delta$ on $\psi_2$, and the Bianchi identities.

The remaining two integrability conditions represent the non-alignment of the Bianchi identities and Maxwell's equations as discussed previously. They can not be directly verified from the commutators and the Bianchi identities. Nevertheless, the gap can be filled in by the Ricci identities. The integrability condition
\be
(\delta'\Delta-\Delta\delta')\log{S}=-\delta'(L_{22}+iL_{23}) - 2i \Delta L_{41},
\ee
after some massages by the Ricci identities, is equivalent to
\begin{equation}
    \delta'(N_{22}-iN_{23})=i(L_{41}-L_{14})(N_{22}-iN_{23}),
\end{equation}
which can be verified from the action of the commutator $\delta' \Delta-\Delta \delta'$ on $\psi_2$ and the Bianchi identities. The last integrability condition 
\be
(\bar{\delta}\delta'-\delta'\bar{\delta})\log{S}=2i \bar{\delta} L_{41} - \frac{1}{\sqrt{2}} \delta' (N_{20}+iN_{30})
\ee
can be verified in a similar way. It is reduced to
\begin{equation}
       \delta'(N_{20}+iN_{30})=(M^2_{34}+iL_{41})(N_{20}+iN_{30}),
\end{equation}
when applying the Ricci identities, which can be proven by the action of the commutator $\bar{\delta}\delta'-\delta'\bar{\delta}$ on $\psi_2$ and the Bianchi identities.

We have proven all the integrability conditions for the scalar equations in \eqref{5dscalar}, which, equivalently, verifies that there must be solutions for the scalar field $S$. Hence, any  type D vacuum solution of this special class in five dimensions admits an algebraically general Maxwell tensor on the type D background that squares to give the Weyl tensor. One can further prove that any solution to the scalar equations in \eqref{5dscalar} must satisfy the wave equation
\begin{equation}
   \left[ \Box - 2\psi_2 + 2(L_{41})^2 \right]S=0.
\end{equation}
A new source term, the spin coefficient $L_{41}$, appears in the wave equation, which is curious since the spin coefficient is gauge-dependent. Nevertheless, it arises from our special reduction of the type D spacetime. We somehow specify the fifth basis $m^4$ as a special direction when imposing $C_{\hat{a}\hat{b}\hat{c}4}=0$, which affects the scalar field. It is amusing to notice that $m^4$ is tangent to a spacelike geodesic. Its deviation tensor can be computed as
\be
\n_\mu m^4_\nu=L_{41}(m^2_\mu m^2_\nu + m^3_\mu m^3_\nu + l_\mu n_\nu + n_\mu l_\nu).
\ee
Clearly, the expansion $L_{41}$ is the only non-zero component in the deviation, which affects the evolution of the scalar $S$ along the fifth direction.

\section{Comments on the solutions lifted from 4d}

The simplest solutions that satisfy the reduction imposed in previous section are those lifted from four-dimensional type D solution by introducing an extra spatial dimension $z$, 
\be
\td s^2_{5D}=\td s^2_{4D} + \td z^2.
\ee
Correspondingly, the fifth vielbein basis is chosen as $m^4=\td z$. However, from the point of view of the WDC, such type of solutions is trivial in the sense that the spin coefficient $L_{41}$ vanishes. The alignment of the Bianchi identities, the Maxwell’s equations, and the wave equation for the scalar field arises as the four-dimensional case. Basically, the five-dimensional WDC formula is a rewriting of the four-dimensional one. For instance, for the five-dimensional rotating black string solution lifted from four-dimensional Kerr solution, the vielbein bases of the solutions in $(t,r,\theta,\phi,z)$ coordinates are given by
\be
\begin{split}
&l=\frac{\Lambda}{2\rho^2} \td t + \frac12 \td r - \frac{a \Lambda \sin^2\theta}{2\rho^2}\td \phi, \\
&n=-\td t + \frac{\rho^2}{\Lambda}td r + a \sin^2\theta \td \phi, \\
&m^2=-\frac{a^2 \cos\theta\sin\theta}{\rho^2}\td t + r \td \theta + \frac{a (r^2 + a^2) \cos\theta\sin\theta}{\rho^2}\td \phi, \\
&m^3= -\frac{a r \sin\theta}{\rho^2}\td t - a \cos\theta \td \theta + \frac{r(r^2+a^2)\sin^2\theta}{\rho^2}\td \phi,
\end{split}
\ee
where $\Lambda=r^2-2 Mr + a^2$, $\rho^2=\bar{\rho}\bar{\rho}^*$, and $\bar\rho=r + i a \cos\theta$. For the black string solution, we obtain
\be
\begin{split}
\psi_2=-\frac{M}{(r-i a\cos\theta)^3},\quad (L_{22}+iL_{23})=\frac{\Lambda}{2\rho^2\bar{\rho}^*},\quad (N_{22}-i N_{23})=\frac{1}{\bar{\rho}^*},\\
\frac{1}{\sqrt{2}}(L_{21}-i L_{31})=-\frac{ia\sin\theta}{\sqrt{2} (\bar\rho^*)^2},\quad \frac{1}{\sqrt{2}}(N_{20} + i N_{30})=\frac{-ia\sin\theta}{\sqrt{2}\rho^2}.
\end{split}
\ee
which are nothing but the Newman-Penrose quantities in four dimensions, see, e.g., in \cite{Chandrasekharkerr}. Hence, the WDC relation for the five-dimensional rotating black string solution is exactly the same as the four-dimensional Kerr solution. However, we want to emphasize that in the case of $L_{41}\neq0$ the five-dimensional type D WDC is significantly distinct from the case for solutions lifted from four dimensions. The verification of the WDC for the case of $L_{41}\neq0$ is rather complicated and tedious. Unfortunately, we do not know any exact solution that satisfies the reduction conditions with $L_{41}\neq0$. We comment on the solutions lifted from four dimensions, e.g., the five-dimensional rotating black string solution, to simply test the WDC relation proposed in previous section. A relevant interesting point for future direction is that one can consider the Kaluza-Klein reduction for type D solutions in five dimensions which have a clear connection to four-dimensional Einstein-Maxwell-dilaton theory. It could provide an alternative higher-dimensional WDC perspective for the four-dimensional sourced WDC \cite{Easson:2021asd}.

\section{Discussions}

In this paper, we revisit the four-dimensional type D WDC relation from the equations of motion perspective. We verify that any four-dimensional type D vacuum solution admits an algebraically general Maxwell field to realize the WDC relation. A scalar field is introduced in the WDC relation which fulfills the wave equation sourced by the only non-zero Weyl scalar. We then extend the type D WDC to five dimensions. We find a self-contained reduction of the type D spacetime in the CMPP classification where a concrete WDC relation is revealed. For the case of two equal electromagnetic fields, the four-dimensional WDC formula constricts that it is only valid for the type D and type N spacetime. In together with a previous work \cite{Zhao:2024ljb}, we have extended both cases to five dimensions.

To close this paper, we will comment on the application of the spinorial formalism for the five-dimensional type D case. We have uncovered the WDC in the vielbein formalism. However, all the other known WDC examples are associated to the spinorial formalism. We have tried to also uncover the spinorial relations. But the only non-zero Weyl scalar arises too many little group bi-spinors in the five-dimensional spinorial formalism \cite{Monteiro:2018xev} to find a concrete WDC relation. Eventually, we did not find any advantage from the spinorial formalism in the understanding of the five-dimensional type D WDC relation.

\section*{Acknowledgments}

The authors are grateful to Hong L\"{u} for useful discussions. This work is supported in part by the National Natural Science Foundation of China (NSFC) under Grants No. 11935009, No. 12375066, and No. 12247103.

\appendix

\section{4d Type D spacetime in the NP formalism}

\subsection{Commutators of the null bases}
\label{4dcommutator}

\begin{align}
        &\delta D-D \delta=(\bar{\alpha}+\beta-\bar{\pi})D-\bar{\rho}\delta,\\
        &\delta \Delta-\Delta \delta=(\tau-\bar{\alpha}-\beta)\Delta+(\mu-\gamma+\bar{\gamma})\delta,\\
        &\bar{\delta}\delta-\delta\bar{\delta}=(\bar{\mu}-\mu)D+(\bar{\rho}-\rho)\Delta+(\alpha-\bar{\beta})\delta+(\beta-\bar{\alpha})\bar{\delta},\\
        &\Delta D-D \Delta=(\gamma+\bar{\gamma})D-(\bar{\tau}+\pi)\delta-(\tau+\bar{\pi})\bar{\delta}.
\end{align}

\subsection{Ricci identities}
\label{4dricci}

\begin{align}
        &D\rho=\rho^2,\\
        &D\tau=\rho(\tau+\bar{\pi}),\\
        &D\alpha=\rho (\alpha+\pi),\\
        &D\beta=\bar{\rho}\beta,\\
        &D\gamma=\alpha(\tau+\bar{\pi})+\beta(\bar{\tau}+\pi)+\tau\pi + \Psi_2,\\
        &D\mu-\delta\pi=\bar{\rho}\mu+\pi(\bar{\pi}+\beta-\bar{\alpha})+\Psi_2,\\
        &\bar{\delta}\pi=-\pi(\pi+\alpha-\beta),\\
        &\Delta\pi=-\mu(\pi+\bar{\tau})-\pi(\gamma-\bar{\gamma}),\\
        &\delta \rho=\rho(\bar{\alpha}+\beta)+\tau(\rho-\bar{\rho}),\\
        &\delta\alpha-\bar{\delta\beta}=\mu \rho+\alpha\bar{\alpha}+\beta\bar{\beta}-2\alpha\beta+\gamma(\rho-\bar{\rho})-\Psi_2,\\
        &\bar{\delta}\mu=-\pi(\mu-\bar{\mu})-\mu(\alpha+\bar{\beta}),\\
        &\Delta\mu=-\mu(\gamma+\bar{\gamma+\mu}),\\
        &\delta\gamma-\Delta\beta=\gamma(\tau-\bar{\alpha}-\beta)+\mu\tau-\beta(\gamma-\bar{\gamma}-\mu),\\
        &\delta\tau=\tau(\tau+\beta-\bar{\alpha}),\\
        &\Delta\rho-\bar{\delta}\tau=-\rho\bar{\mu}+\tau(\bar{\beta}-\alpha-\bar{\tau})+\rho(\gamma+\bar{\gamma})-\Psi_2,\\
        &\Delta\alpha-\bar{\delta}\gamma=\alpha(\bar{\gamma}-\bar{\mu})+\gamma(\bar{\beta}-\bar{\tau}).
\end{align}

\section{Reductions of the type D spacetime in 5d}
\label{5dreduction}

We present the complete constraints from the type D conditions and the reduction imposed in the main text. The trace-free property of the Weyl tensor requires
\begin{equation}
    C_{0\hat{b}\hat{c}1}+C_{1\hat{b}\hat{c}0}+C_{2\hat{b}\hat{c}2}+C_{3\hat{b}\hat{c}3}=0,
\end{equation}
which yields the following relations
\begin{equation}
 C_{0331}=C_{0221},\quad C_{0101}=-2C_{0221},\quad C_{2323}=2C_{0221},
        \quad C_{0321}=-C_{0231}.
\end{equation}
The cyclic identity yields
\begin{equation}
    C_{0123}=-2C_{0231}.
\end{equation}
The five dimensional Goldberg-Sachs-like theorem \cite{Durkee:2009nm,Ortaggio:2012hc} ensures that $L_{10}=L_{i0}=0$ after choosing the Weyl aligned null directions as the principle null directions. Then, one can set $M^i_{\ j0}=0$ by performing an appropriate spin transformation. 

Now, we consider the Bianchi identities that involve the evolution of the vanishing components from the type D condition. This class of equations include equations (B.8), (B.10), (B.14), and (B.15) of \cite{Pravda:2004ka}. With our choice of gauge, (B.8) is automatically satisfied. (B.10), (B.14) and (B,15) give the following constraints for spin coefficients
\begin{equation}
    \begin{split}
        &C_{01jk}N_{i1}+C_{0j1i}N_{k1}-C_{0k1i}N_{j1}-C_{iljk}N_{l1}=0,\\
        &C_{01\{ jm|}N_{i|k\}}-C_{0\{ j|1i}N_{|km\}}+C_{0\{j|1i}N_{|mk\}}+C_{il\{jk|}N_{l|m\}}=0,\\
        &C_{01\{jk|}L_{i|m\}}-C_{0i1\{j}L_{km\}}+C_{0i1\{j}L_{mk\}}+C_{il\{jk|}L_{l|m\}}=0.
    \end{split}
\end{equation}
The above algebraic equations yield
\begin{equation}
    \begin{split}
    &N_{i1}=0,\\
    &N_{43}=3N_{34},\quad N_{42}=3N_{24}, \quad  N_{44}=0,\\
    &L_{43}=3L_{34},\quad L_{42}=3L_{24}, \quad L_{44}=0.
 \end{split}
\end{equation}
Then, we consider the rest of the Bianchi identities. By comparing the coefficients of repeated equations within these classes, we can obtain additional constraints on the spin coefficients. All the non-trivial relations from (B.1) are as follows
\begin{equation}
    \begin{split}
        &2\delta_2C_{0221}=3C_{0221}(L_{21}+N_{20})+3C_{0231}(L_{31}-N_{30}),\\
        &2\delta_3C_{0221}=3C_{0221}(L_{31}+N_{30})+3C_{0231}(N_{20}-L_{21}),\\
        &\delta_4C_{0221}=C_{0221}(L_{41}+N_{40}).\\
    \end{split}
\end{equation}
All the non-trivial relations from (B.2) are as follows
\begin{equation}
    \begin{split}
        &C_{0221}N_{42}+C_{0231}(N_{43}+2M^3_{41})=0,\\
        &C_{0221}N_{43}-C_{0231}(N_{42}+2M^2_{41})=0,\\
        &2\Delta C_{0231}=3C_{0221}(N_{23}-N_{32})-3C_{0231}(N_{22}+N_{33}).
    \end{split}
\end{equation}
All the non-trivial relations from (B.3) are as follows
\begin{equation}
    \begin{split}
        &L_{34}=L_{43},\\
        &L_{24}=L_{42},\\
        &2DC_{0231}=-3C_{0221}(L_{23}-L_{32})-3C_{0231}(L_{22}+L_{33}).
    \end{split}
\end{equation}
Combining the results from (B.3) and (B.15) we have
\begin{equation}
    L_{24}=L_{42}=L_{34}=L_{43}=0.
\end{equation}
When the indices include the fifth direction, (B.4) gives the following two pairs of constraints 
\begin{equation}
    \begin{split}
        &C_{0221}(2N_{42}+M^2_{41})+C_{0231}M^3_{41}=0,\\
        &C_{0221}(2N_{43}+M^3_{41})-C_{0231}M^2_{41}=0,
    \end{split}
\end{equation}
and
\begin{equation}
    \begin{split}
        &C_{0221}(-N_{42}+M^2_{41})+C_{0231}(N_{43}-M^3_{41})=0,\\
        &C_{0221}(-N_{43}+M^3_{41})+C_{0231}(-N_{42}+M^2_{41})=0.
    \end{split}
\end{equation}
As we assume that $C_{0221},C_{0231}$ are non-zero, the above equations yield
\begin{equation}
    N_{43}=N_{34}=N_{42}=N_{24}=M^2_{41}=M^3_{41}=0.
\end{equation}
The remaining relations from (B.4) are as follows
\begin{equation}
    \begin{split}
        &\Delta C_{0221}=-3C_{0221}N_{22}+3C_{0231}N_{32},\\
        &\Delta C_{0331}=-3C_{0221}N_{33}-3C_{0231}N_{23}.
    \end{split}
\end{equation}
Hence
\begin{equation}
    N_{22}=N_{33},\quad N_{32}=-N_{23},
\end{equation}
and 
\begin{equation}
    \Delta C_{0231}=3C_{0221}N_{23}-3C_{0231}N_{22},
\end{equation}
which is equivalent to the last equation from (B.2). We continue with (B.5) which yields
\begin{equation}
    \begin{split}
        &L_{22}=L_{33},\quad L_{32}=-L_{23},\\
        &DC_{0221}=-3C_{0221}L_{22}+3C_{0231}L_{23},\\
        &DC_{0231}=-3C_{0231}L_{22}-3C_{0221}L_{23},
    \end{split}
\end{equation}
where the last one is equivalent to the last relation from (B.3). Next, (B.6) gives that
\begin{equation}
    \begin{split}
        &M^2_{42}=M^{3}_{43},\quad M^3_{42}=-M^2_{43},\quad M^2_{44}=M^3_{44}=0,\quad L_{41}+N_{40}=2M^2_{42},\\
        &C_{0221}(L_{41}-N_{40})-2C_{0231}M^2_{43}=0,\\
        &2\delta_2C_{0231}=3C_{0221}(N_{30}-L_{31})+3C_{0231}(L_{21}+N_{20}),\\
        &2\delta_3C_{0231}=3C_{0221}(L_{21}-N_{20})+3C_{0231}(L_{31}+N_{30}),\\
        &\delta_4C_{0231}=C_{0231}(L_{41}+N_{40}).
    \end{split}
\end{equation}
When the indices include the fifth direction, (B.7) gives that
\begin{equation}
\begin{split}
        &C_{0221}M^2_{43}+C_{0231}(L_{41}-M^2_{42})=0,\\
        &\delta_4C_{0221}=C_{0221}(L_{41}+M^2_{42})+C_{0231}M^3_{42}.
\end{split}
    \end{equation}
Comparing them with (B.1) and (B.6), we obtain
\begin{equation}
    L_{41}=N_{40}=M^2_{42}=M^3_{43},\quad M^3_{42}=M^2_{43}=0.
\end{equation}
With the above constraints on the spin coefficients, one can verify that the remaining Bianchi identities (B.7), (B.9), (B.12), (B.13) and (B.16) are redundant equations that do not lead to new constraint.

\section{5d  reduced Type D spacetime in the vielbein formalism}

\subsection{Bianchi identities}
\label{Bianchi5d}

\begin{align}
        &DC_{0221}=-3C_{0221}L_{22}+3C_{0231}L_{23},\\
        &DC_{0231}=-3C_{0231}L_{22}-3C_{0221}L_{23},\\
        &\Delta C_{0221}=-3C_{0221}N_{22}-3C_{0231}N_{23},\\
        &\Delta C_{0231}=3C_{0221}N_{23}-3C_{0231}N_{22},\\
        &2\delta_2C_{0231}=3C_{0221}(N_{30}-L_{31})+3C_{0231}(L_{21}+N_{20}),\\
        &2\delta_3C_{0231}=3C_{0221}(L_{21}-N_{20})+3C_{0231}(L_{31}+N_{30}),\\
        &2\delta_2C_{0221}=3C_{0221}(L_{21}+N_{20})+3C_{0231}(L_{31}-N_{30}),\\
        &2\delta_3C_{0221}=3C_{0221}(L_{31}+N_{30})+3C_{0231}(N_{20}-L_{21}),\\
        &\delta_4C_{0221}=2C_{0221}L_{41},\\
        &\delta_4C_{0231}=2C_{0231}L_{41}.
\end{align}

\subsection{Maxwell's equations}
\label{Maxwell5d}

\begin{align}
        &DF_{01}=-2F_{01}L_{22}+2F_{23}L_{23},\\
         &DF_{23}=-2F_{01}L_{23}-2F_{23}L_{22},\\
        &\Delta F_{01}=-2F_{01}N_{22}-2F_{23}N_{23},\\
        &\Delta F_{23}=2F_{01}N_{23}-2F_{23}N_{22},\\
        &\delta_2F_{01}=F_{01}(L_{21}+N_{20})+F_{23}(L_{31}-N_{30}),\\
         &\delta_2F_{23}=F_{01}(L_{31}-N_{30})+F_{23}(L_{21}+N_{20}),\\
        &\delta_3F_{23}=F_{01}(L_{21}-N_{20})+F_{23}(L_{31}+N_{30}),\\
        &\delta_3F_{01}=F_{01}(L_{31}+N_{30})-F_{23}(L_{21}-N_{20}),\\
        &\delta_4F_{01}=2F_{01}L_{41},\\        
        &\delta_4F_{23}=2F_{23}L_{41}.
\end{align}

\subsection{Commutators of the basis vectors}
\label{commutator5d}
We present the commutators of the derivatives we defined in the main text
\begin{align}
&\delta D - D \delta  =\frac{1}{\sqrt{2}}[(L_{12}+N_{20})-i(L_{13}+N_{30})]D+(L_{22}-iL_{23})\delta, \\
&\bar{\delta}\Delta-\Delta \bar{\delta}=\frac{1}{\sqrt{2}}(N_{21}+iN_{31})D+\frac{1}{\sqrt{2}}[(L_{21}-L_{12})+i(L_{31}-L_{13})]\Delta \nn \\
&\hspace{3cm}+[N_{22}+i(N_{23}+M^2_{31})]\bar{\delta},\\
&\delta' D -D \delta' =i(L_{14}+N_{40})D, \\
& \delta \delta' - \delta'\delta =(M^2_{34}-iL_{41})\delta,\\
&\Delta D -D \Delta =L_{11}D+\frac{1}{\sqrt{2}}[(L_{21}-N_{20})+i(L{31}-N_{30})]\delta, \\
&\bar{\delta}\delta-\delta\bar{\delta}=i(N_{32}-N_{23})D+i(L_{32}-L_{23})\Delta-\frac{1}{\sqrt{2}}(M^2_{33}-iM^2_{32})\delta \nn \\
&\hspace{3cm}+\frac{1}{\sqrt{2}}(M^2_{33}+iM^2_{32})\bar{\delta},\\
& \bar{\delta}D-D\bar{\delta}=\frac{1}{\sqrt{2}}[(L_{12}+N_{20})+i(L_{13}+N_{30})]D+(L_{22}+iL_{23})\bar{\delta}, \\
&\delta \Delta-\Delta \delta=\frac{1}{\sqrt{2}}(N_{21}-iN_{31})D+\frac{1}{\sqrt{2}}[(L_{21}-L_{12})-i(L_{31}-L_{13})]\Delta \nn \\
&\hspace{3cm}+(N_{22}-iN_{23}-iM^2_{31})\delta,\\
&  \delta' \Delta-  \Delta \delta' =iN_{41}D+i(L_{41}-L_{14})\Delta, \\
&\bar{\delta}\delta'-\delta'\bar{\delta}=-(M^2_{34}+iL_{41})\bar{\delta}.
\end{align}

\subsection{Useful Ricci identities}
\label{ricci5d}

Here we list the components of the Ricci identity that we have applied in order to verify the integrability conditions.
\begin{equation}
\begin{split}
    &DL_{21}=L_{22}(-L_{21}+N_{20})+L_{23}(-L_{31}+N_{30}),\\
    &DL_{31}=L_{22}(-L_{31}+N_{30})-L_{23}(-L_{21}+N_{20}),
\end{split}
\end{equation}
and
\begin{equation}
    \begin{split}
        &\delta_2L_{22}+\delta_3L_{23}=L_{12}L_{22}+(L_{13}+2L_{31})L_{23},\\
        &\delta_3L_{22}-\delta_2L_{23}=L_{13}L_{22}-(L_{12}+2L_{21})L_{23},
    \end{split}
\end{equation}
are used to prove the first integrability condition.
\begin{equation}
    \begin{split}
        &\Delta N_{20}=N_{22}(-N_{20}+L_{21})+N_{23}(-N_{30}+L_{31})+N_{30}M^2_{31},\\
        &\Delta N_{30}=-N_{23}(-N_{20}+L_{21})+N_{22}(-N_{30}+L_{31})-N_{20}M^2_{31},
    \end{split}
\end{equation}
and
\begin{equation}
    \begin{split}
        &\delta_2N_{22}+\delta_3N_{23}=-L_{12}N_{22}+(2N_{30}-L_{13})N_{23},\\
        &\delta_3N_{22}-\delta_2N_{23}=-L_{13}N_{22}+(L_{12}-2N_{20})N_{23},
    \end{split}
\end{equation}
are used to prove the second integrability condition.
\begin{equation}
    DL_{41}=0,
\end{equation}
and
\begin{equation}
    \begin{split}
        &\delta_4L_{22}=L_{22}(L_{14}+L_{41}),\\
        &\delta_4L_{23}=L_{23}(L_{14}+L_{41}),        
    \end{split}
\end{equation}
are used to prove the third integrability condition.
\begin{equation}
    \begin{split}
        &\delta_2L_{41}=\delta_3L_{41}=0,\\
        &\delta_4L_{21}=L_{21}L_{41}+L_{31}M^2_{34},\\
        &\delta_4L_{31}=L_{31}L_{41}-L_{21}M^2_{34},
    \end{split}
\end{equation}
are used to prove the forth integrability condition.
\begin{equation}
    \begin{split}
        &\Delta L_{22}-\delta_2L_{21}=L_{11}L_{22}-L_{21}L_{21}-L_{31}M^2_{32}-L_{22}N_{22}+L_{23}N_{23}+C_{0221},\\
        &\Delta L_{22}-\delta_3L_{31}=L_{11}L_{22}-L_{31}L_{31}+L_{21}M^2_{33}+L_{23}N_{23}-L_{22}N_{22}+C_{0221},\\
        &\Delta L_{23}-\delta_3L_{21}=L_{11}L_{23}-L_{21}L_{31}-L_{31}M^2_{33}-L_{22}N_{23}-L_{23}N_{22}+C_{0231},\\
        &\Delta L_{23}+\delta_2 L_{31}=L_{11}L_{23}+L_{21}L_{31}-L_{21}M^2_{32}-L_{23}N_{22}-L_{22}N_{23}+C_{0231},
    \end{split}
\end{equation}
and
\begin{equation}
    \begin{split}
        &DN_{22}-\delta_2N_{20}=-N_{20}N_{20}-N_{30}M^2_{32}-N_{22}L_{22}+N_{23}L_{23}+C_{0221},\\
        &DN_{22}-\delta_3N_{30}=-N_{30}N_{30}+N_{20}M^2_{33}+N_{23}L_{23}-N_{22}L_{22}+C_{0221},\\
        &DN_{23}-\delta_3N_{20}=-N_{20}N_{30}-N_{30}M^2_{33}-N_{22}L_{23}-N_{23}L_{22}+C_{0231},\\
        &DN_{23}+\delta_2N_{30}=N_{20}N_{30}-N_{20}M^2_{32}-N_{23}L_{22}-N_{22}L_{23}-C_{0231},
    \end{split}
\end{equation}
are used to prove that the fifth and sixth integrability conditions are equivalent.
\begin{equation}
    \begin{split}
        &\delta_4N_{22}=N_{22}(L_{41}-L_{14}),\\
       &\delta_4N_{23}=N_{23}(L_{41}-L_{14}),\\
       &\Delta L_{41}=0.
   \end{split}
\end{equation}
and
\begin{equation}
    \begin{split}
&\delta_4N_{20}=N_{20}L_{41}+N_{30}M^2_{34},\\
        &\delta_4N_{30}=N_{30}L_{41}-N_{20}M^2_{34},\\
        &\bar{\delta}L_{41}=0.
           \end{split}
\end{equation}
are used to prove the last two integrability conditions.

\providecommand{\href}[2]{#2}\begingroup\raggedright\endgroup

\end{document}